\documentclass[a4paper,twocolumn,11pt]{quantumarticle}
\pdfoutput=1
\usepackage[utf8]{inputenc}
\usepackage[english]{babel}
\usepackage[T1]{fontenc}
\usepackage[numbers,super]{natbib}
\usepackage{amsmath,amssymb,amsfonts}
\usepackage{listings}

\usepackage{hyperref}
\usepackage{algpseudocode}
\usepackage{algorithm}
\usepackage{graphicx}
\usepackage{textcomp}
\usepackage{xcolor}

\usepackage{booktabs}
\usepackage{braket}

\newcommand{\norm}[1]{\left\lVert#1\right\rVert}

\begin{document}

\title{Quantum Data Reduction with Application to Video Classification}

\author{Kostas Blekos}
\affiliation{{Computer Engineering \& Informatics Dept.}, {University of Patras}, Rio, Greece}
\email{mplekos@upatras.gr}
\orcid{0000-0002-6777-2107}
\author{Dimitrios Kosmopoulos}
\affiliation{{Computer Engineering \& Informatics Dept.}, {University of Patras}, Rio, Greece}
\email{dkosmo@upatras.gr}

\maketitle

\begin{abstract}
We investigate a quantum video classification method using a hybrid algorithm.
A quantum-classical step performs a data reduction on the video dataset and a
quantum step---which only has access to the reduced dataset---classifies the
video to one of $k$ classes. We verify the method using sign videos and
demonstrate that the reduced dataset contains enough information to
successfully classify the data, using a quantum classification process.

The proposed data reduction method showcases a way to alleviate the ``data loading'' problem of quantum computers
for the problem of video classification.
Data loading is a huge bottleneck, as there are no known
efficient techniques to perform that task without sacrificing many of the benefits of quantum computing.

\end{abstract}


\section{Introduction}

In the past few years there has been a lot of interest in exploring quantum
computing applications for classical computational problems.
Quantum computing, by taking advantage of the quantum mechanical properties of nature,
offers a new toolbox for attacking such problems with potentially great benefits
in efficiency and performance.
Image and video classification algorithms are an important subset of algorithms
where quantum computing applications are actively
investigated~\cite{dangImageClassificationBased2018,zhouQuantumKNearestNeighborImage2021,schuldCircuitcentricQuantumClassifiers2020,ostaszewskiQuantumImageClassification2015,farhiClassificationQuantumNeural2020,hurQuantumConvolutionalNeural2021,lloydQuantumEmbeddingsMachine2020},
though, to our best knowledge,
applications on video processing are only sparsely researched~\cite{blekosQuantum3DConvolutional2021}.

One important issue with quantum video processing applications is that they require
the transfer of large amount of data
from the classical machines
to the quantum computer.
This ``data loading'' of the quantum computer
is a huge bottleneck as there are no known
efficient techniques
that can perform such a task without sacrificing many of the quantum-gains~\cite{harrowSmallQuantumComputers2020,haugLargescaleQuantumMachine2021,liangVariationalQuantumAlgorithms2020,corteseLoadingClassicalData2018}.
This problem is not unique to video processing applications~\cite{biamonteQuantumMachineLearning2017}.
For example, solving a system of linear equations through the HHL algorithm is usually
stated as needing $O(\log(n))$ steps for $n$ equations.
This statement, though, hides many assumptions, one of which is that one should be able to read and store $n$
parameters on at most $O(\log(n))$ steps~\cite{aaronsonReadFinePrint2015}.

In general,
it is possible that near-term applications will mostly concern the cooperation
of the classical hardware with small quantum computing units~\cite{harrowSmallQuantumComputers2020}. 
This highlights
the importance of reducing the classical data space before communicating with
the quantum hardware.
One proposal, by Harrow~\cite{harrowSmallQuantumComputers2020}, is using hybrid classical-quantum
algorithms where a ``representative'' subset of a problem's data set is extracted.
This subset is then used as input to the quantum computer instead of the whole data set.

\subsection{Contribution}
In this work we propose a data reduction scheme on a hybrid classical-quantum algorithm for video classification. 
Using a hybrid procedure, we extract the most important pixels of a video which we then use as the reduced ``training set''.
We classify each new video based only on these pixel distributions.
We perform simulations of the proposed algorithm
verifying that meaningful information can be extracted from just a small percentage of the initial video.
The videos are accurately classified showing that the algorithm is
capable of accurately identifying the most significant
pixels using an efficient quantum procedure.
To the best of our knowledge this is the first such approach.

In Section~\ref{sec:back} we provide the background knowledge. In Section~\ref{sec:method} we describe the proposed method. In Section~\ref{sec:results} we present our results on a public dataset. 
Finally in Section~\ref{sec:discuss} we discuss the merits and constraints of the method.

\section{Quantum background}%
\label{sec:back}
Quantum computers 
use quantum states of two levels (a \emph{qubit}) to store and process information
instead of using bits of 0 and 1.
Abstractly, a qubit
is a two-dimensional vector 
of complex parameters and norm 1, i.e.,
$\ket{\text{qubit}} = \begin{pmatrix}a\\b \end{pmatrix}$ with $a,b\in\mathbb{C}$
and $|a|^2 + |b|^2 = 1$.
Similarly, 
a series of $n$ quantum bits (often referred to as a \emph{quantum register})
form a $2^n$-dimensional vector of complex parameters and norm 1.
We define
the \emph{computational basis} as the one-hot orthonormal basis:
\begin{align*}
\ket{0} = \begin{pmatrix}1 \\0 \\\ldots\\ 0\end{pmatrix}, 
\ket{1} = \begin{pmatrix}0 \\1 \\ \ldots\\ 0\end{pmatrix}, 
\ldots
\ket{2^n-1} = \begin{pmatrix}0 \\0 \\ \ldots\\ 1\end{pmatrix}, 
\end{align*}
Using these definitions, a general quantum register is written
as a linear combination of the computational basis
$\ket{q} = a_0\ket{0} + a_1\ket{1} + \cdots + a_{2^n-1}\ket{2^n-1}$.
The parameter $a_i$ is called the \emph{amplitude} of the state $\ket{i}$.
The actual parameters of a quantum register are unknowable.
When the value of a quantum register is needed
a \emph{measurement} is performed.
The result of the measurement
gives \emph{one} of the computational basis vectors.
In particular, the result of measuring the register $\ket{q}$ above
is $\ket{i}$ with probability $|a_i|^2$.

Quantum registers are manipulated by use of quantum gates (also called quantum operators)
which act as the quantum analog of the classical logical gates (AND, OR, NOT, etc).
Quantum gates are represented by complex unitary matrices
of appropriate dimensions and their action is calculated by
simple matrix multiplication.

\subsection{Inner product estimation}\label{sec:innerp}

The \emph{Inner product estimation} is an efficient subroutine for the estimation
of the inner product between two quantum registers $\ket{a}, \ket{b}$.
The subroutine has three steps.
Firstly an ancilla qubit is entangled with the two quantum registers 
producing the register:
\begin{align*}
    \ket{\phi} &= \frac1{\sqrt{2}}\Bigl(\ket{0}_{\alpha}\ket{a} + \ket{1}_{\alpha}\ket{b}\Bigr)\\
\end{align*}
Secondly, a Hadamard gate $H = \frac1{\sqrt{2}}\begin{pmatrix}1&1\\1&-1\end{pmatrix}$
with $H\ket{0} = \frac1{\sqrt{2}}\Bigl(\ket{0} + \ket{1}\Bigr)$ and
$H\ket{1} = \frac1{\sqrt{2}}\Bigl(\ket{0} - \ket{1}\Bigr)$,
is applied on the ancilla qubit:
\begin{align*}
    H_{\alpha}\ket{\phi} &= \frac1{\sqrt{2}}\Biggl(\frac1{\sqrt{2}}\Bigl(\ket{0}_{\alpha}+\ket{1}_{\alpha}\Bigr)\ket{a}\\
			 &+ \frac1{\sqrt{2}}\Bigl(\ket{0}_{\alpha}-\ket{1}_{\alpha}\Bigr)\ket{b}\Biggr)\\
			 &= \frac12\left(\ket{0}_{\alpha}\Bigl(\ket{a} + \ket{b}\Bigr) + \ket{1}_{\alpha}\Bigl(\ket{a} - \ket{b}\Bigr)\right)
\end{align*}
Then, 
a measurement of the ancilla qubit is performed.
The probability of measuring $\ket{0}_{\alpha}$ is its amplitude squared, thus:
\begin{align*}
    P(\ket{0}_{\alpha}) &= \norm{\frac12\left(\ket{a} + \ket{b}\right)}^2\\
		&= \frac14\sqrt{\sum_i (\overline{a_i +b_i})(a_i+b_i)}^2\\
		&= \frac14\left|\sum_i \overline{a}_ia_i + \overline{a}_ib_i + \overline{b}_ia_i + \overline{b}_ib_i \right|\\
		&= \frac14\left| \braket{a|a} + \braket{a|b} + \braket{b|a} + \braket{b|b} \right|\\
		&= \frac{1 + \braket{a|b}}{2}
\end{align*}
since, by definition, $\braket{\cdot|\cdot} \geq 0$ and $\braket{x|x}=1$.
Therefore, by repeatedly measuring the ancilla qubit
an estimate for the inner product $\braket{a|b}$ can be calculated.

\subsection{Amplitude encoding}\label{sec:amplitude}
One often employed method to encode classical data
to quantum registers is the \emph{amplitude encoding}
in which one encodes the information of a classical vector
on the amplitudes of the computational basis.
So, a vector $v = \begin{pmatrix} v_0, & v_1, & \ldots, & v_n\end{pmatrix}$
is encoded to the quantum register
\begin{equation*}
    \ket{v} = \frac 1{\lVert v\rVert}\sum v_i\ket{i}.
\end{equation*}

\subsection{QRAM}\label{sec:qram}
A ``Quantum RAM'' (QRAM) is a classical or quantum data structure that outputs quantum states.
Like its classical analog RAM, it is used to store and retrieve information in a ``Random Access'' model, i.e.\ any desired bit of information
can be addressed individually at will.
Various QRAM models have been proposed that provide efficient
implementations for the crucial quantum state storing and retrieving
procedures~\cite{kerenidisQuantumRecommendationSystems2016,giovannettiQuantumRandomAccess2008}
The inner product estimation algorithm, for example,
that is used in this work, 
can be efficiently implemented using a QRAM~\cite{kerenidisQmeansQuantumAlgorithm2018}.

\section{Methodology}%
\label{sec:method}
We are going to demonstrate the proposed method by means of a video classification application.
The task is to classify short videos displaying different types of hand motion into one of $k$ classes. The methodology can be easily generalized to classification tasks involving high-volume data.

\begin{figure*}[hp]
    \centering
    \includegraphics[width=0.9\textwidth]{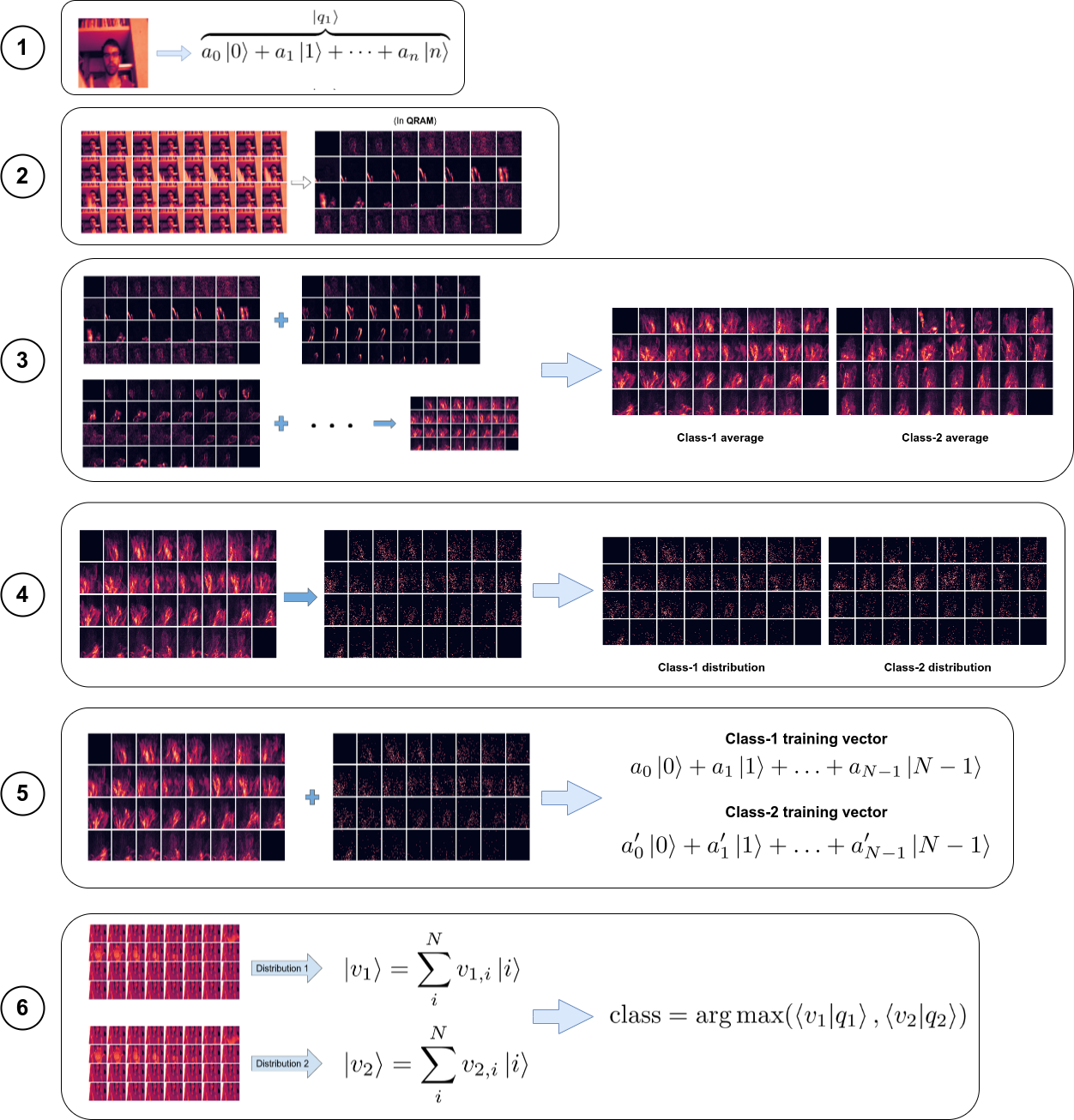}
    \captionsetup{singlelinecheck=off}
    \caption[Schematic of the algorithm.]{%
	Schematic of the algorithm.
	\begin{enumerate}
	    \item Convert training set to quantum states.
	    \item Transform and store in a QRAM\@.
	    \item Repeat (1--2) and keep average of each class.
	    \item Perform $2^q$ measurements on the averages of each class.
	    \item Use measurement results to produce a training quantum register for each class.
	    \item Perform inner product estimation. Classify to best match.
	\end{enumerate}
    }\label{fig:alg}
\end{figure*}

We assume that each video is represented by an $N \times N \times T$ matrix (height$\times$width$\times$frames) of
values in the range [0..1] and belongs to one of $k$ classes. 
We will reduce the information contained in this
$N\times N\times T$-sized matrix to a $2^q$-sized quantum state  using $q$ qubits.
When a classical-to-quantum video conversion is needed
we use the amplitude encoding as described in Section~\ref{sec:amplitude}.
A training set consists of $M\times k$ videos with $M$ videos for each class.
The overview of the algorithm is given in Algorithm~\ref{alg:alg1}
and schematically in Figure~\ref{fig:alg}.

\textbf{1. Convert frames of the training set to quantum states.}
Using the amplitude encoding we convert successive frames to quantum states.

\textbf{2. Perform ``difference'' transform on the quantum states and store in a QRAM\@.}
A \emph{difference transform} efficiently converts two successive frames
$\ket{q_1}$ and $\ket{q_2}$ to their difference $\ket{q_1} - \ket{q_2}$ by
using an ancilla qubit and a Hadamard gate, similarly to the inner product estimation
subroutine of Section~\ref{sec:innerp}:
\begin{align*}
    \ket{\phi} &= \frac1{\sqrt{2}}\Bigl(\ket{0}_{\alpha}\ket{q_1} + \ket{1}_{\alpha}\ket{q_2}\Bigr) \\
    H_{\alpha}\ket{\phi}&=\frac1{\sqrt{2}}\left(\ket{0}_{\alpha}\Bigl(\ket{q_1} + \ket{q_2}\Bigr) 
    + \ket{1}_{\alpha}\Bigl(\ket{q_1} - \ket{q_2}\Bigr)\right) \\
\end{align*}
By measuring the ancilla until we find it at the state $\ket{1}_{\alpha}$
we store the difference $\ket{q_1} - \ket{q_2}$ in the QRAM\@.

\textbf{3. Repeat (1--2) for all the frames of each video averaging over all training videos for each class.}
We repeat the previous steps until all difference frames of all training videos are stored in the QRAM\@.
An average difference-video is then calculated for each class: ${\text{diff-video}_1}, \ldots, {\text{diff-video}_k}$.

\textbf{4. Load average videos and perform $2^q$ measurements to get a distribution for each class.}
We load the $k$ difference videos to quantum registers of size $N\times N\times T$.
We measure $2^q$ times to get a distribution---with replacement---that represents
the most important pixels from all frames, as these are the pixels that have the highest
amplitudes.
This creates a \emph{weighted} distribution of the most important pixels for each class of videos.
Since the amplitudes have been calculated by the differences between 
successive frames, a low amplitude means that there is no significant change
between the frames whereas a high amplitude means that there is significant
difference between the successive frames.
At the end of this procedure we have $k$ distributions of $2^q$ pixel-coordinates.

\textbf{5. Use the distributions and the average videos to produce a training quantum register for each class.}
Using the class distributions we convert each average video to a quantum register using $q$ qubits.
All pixels not belonging to the distributions are ignored.
We produce $k$ training quantum registers $\ket{v_1}, \ldots, \ket{v_k}$.

\textbf{6. Classify a new video by reducing it using the distributions and performing an inner product estimation.}
To classify a new video we use the class distributions to convert it to a quantum register of $q$ qubits.
We keep as amplitudes of the video the pixels that belong to the class distributions and ignore all other pixels. 
For each video we produce $k$ test quantum registers $\ket{V_1}, \ldots, \ket{V_k}$.
We use the inner product estimation subroutine to calculate the $k$ inner products:
$\braket{v_1|V_1}, \ldots, \braket{v_k|V_k}$.
We assign the video to the class:
\begin{equation*}
    \text{class} = \arg\max\Bigl(\braket{v_1|V_1}, \ldots, \braket{v_k|V_k}\Bigr).
\end{equation*}

\textbf{Effect of the number of qubits in performance.} 
The number of qubits $q$ used in the algorithm affects the performance
of the sampling steps (4, 5) and the classification step (6).
Performances of steps 1, 2 and 3 depend dominantly on the dimensions of the dataset.
It is a basic assumption of this work, as stated earlier,
that converting classical data to quantum data is an expensive process
that we need to avoid.
We, therefore, off-load the data conversion to few key instances of the training steps
and then use a reduced dataset to make the prediction.
In other words, we convert the performance of step 6---the data-loading-depended prediction
step---from
$O(f(N\times N\times t)g(N\times N\times t))$
to
$O(f(2^q)g(2^q))$, 
where $f(n)$ is some function that models
the performance of the data-loading procedure and
$g(n)$ is some function that models the performance of the prediction procedure.
At the moment it seems that a relation of
$O(f(n)) \sim O(n)$ is unavoidable~\cite{harrowSmallQuantumComputers2020}.
On the other hand, the simplistic $g$ that we use here might have
$O(g(n)) \sim O(\sqrt{\log(n)})$~\cite{anshuDistributedQuantumInner2021,odonnellEfficientQuantumTomography2016}.

\begin{algorithm}[H]
\caption{Video classification through quantum data reduction}%
\label{alg:alg1}
\begin{algorithmic}
\For{All videos}
\State{}Convert frames to quantum registers.
\State{}Apply ``difference'' transform to consecutive frames.
\State{}Store result in QRAM\@.
\EndFor
\For{All classes}
\State{}Calculate average video.
\State{}Get sample distribution by measuring $2^q$ times each average video.
\State{}Get training quantum register by sampling the average video using the class distribution.
\EndFor
\State{}Reduce a new video by sampling using the class distributions.
\State{}Classify by taking the inner product estimation and setting
    $\text{class} = \arg\max\Bigl(\braket{v_1|V_1}, \ldots, \braket{v_k|V_k}\Bigr)$.
\end{algorithmic}
\end{algorithm}

\section{Experimental Results}%
\label{sec:results}

\begin{figure}[!h]
    \centering
    \includegraphics[width=0.95\columnwidth]{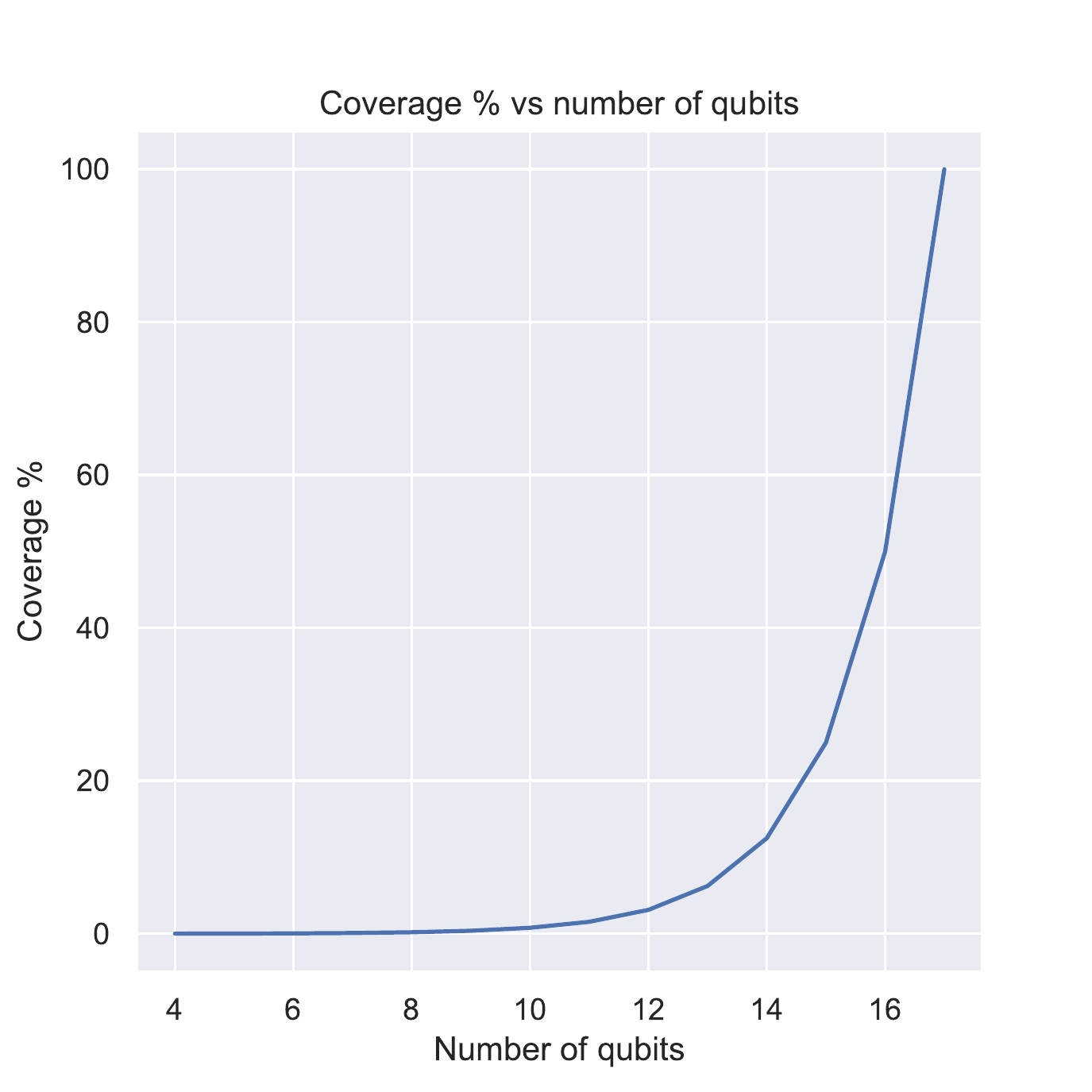}
    \caption{Percentage of the initial video that a quantum register of $q$ qubits encodes.}\label{fig:cover}
\end{figure}

\textbf{Dataset and pre-processing}
We evaluate the proposed algorithm using a small subset
of the ``20BN-jester Dataset V1'' that contains
labeled video clips showing humans performing predefined hand gestures~\cite{materzynskaJesterDatasetLargescale2019}.
We crop and downscale each video so that all frames are 64$\times$64 pixels and all videos 32 frames long $N=64,T=32$.
We perform the simulations using two 
(``Swiping Left'', ``Pulling Hand In''),
three (``Swiping Left'', ``Pulling Hand In'', ``Pushing Hand Away'') and
four (``Swiping Left'', ``Pulling Hand In'', ``Pushing Hand Away'', ``Swiping Right'')
of the available classes ($k=2,3,4$).
We used training sets of sizes $M\times k$ for values $M=20, 40, 60, 80, 120$.
We encode the reduced data using $q$ qubits for values $q=[4..17]$.
We stress that the actual data reduction is logarithmic with the number of qubits
as shown in Figure~\ref{fig:cover}.
Encoding a video using $q=4$ qubits corresponds to a $(64*64*32) / 2^4 = 8192$ reduction.

\textbf{Simulation}
For each combination of the parameters (training set size, number of classes, number of qubits that encode the reduced data)
we simulated the quantum procedures.
We run the simulations for at least 100 iterations
and averaged out the results.

\begin{figure}[!h]
    \centering
    \includegraphics[width=0.95\columnwidth]{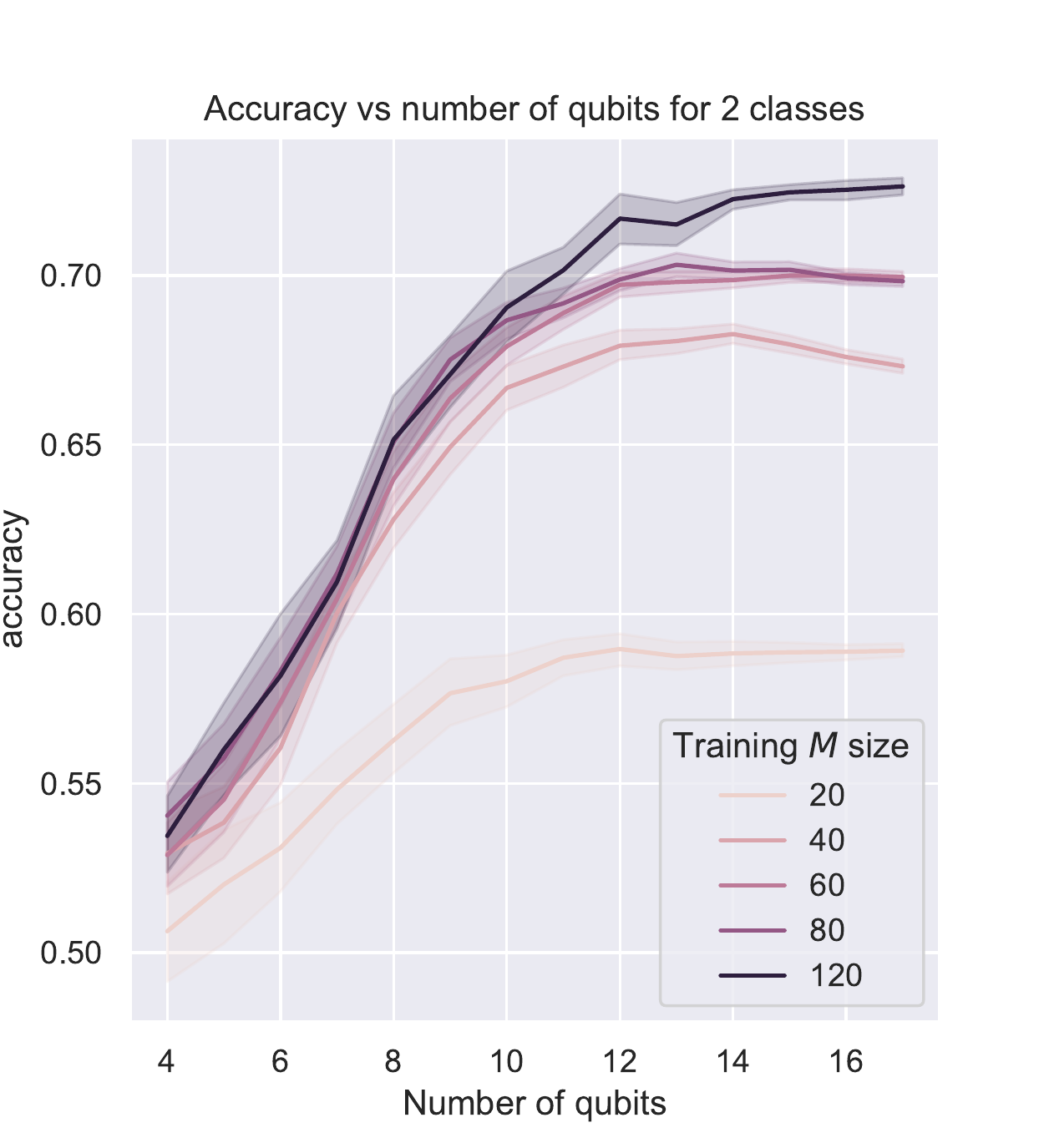}
    \caption{Accuracy vs number of qubits for 2 classes}\label{fig:acc2}
\end{figure}

\begin{figure}[!h]
    \centering
    \includegraphics[width=0.95\columnwidth]{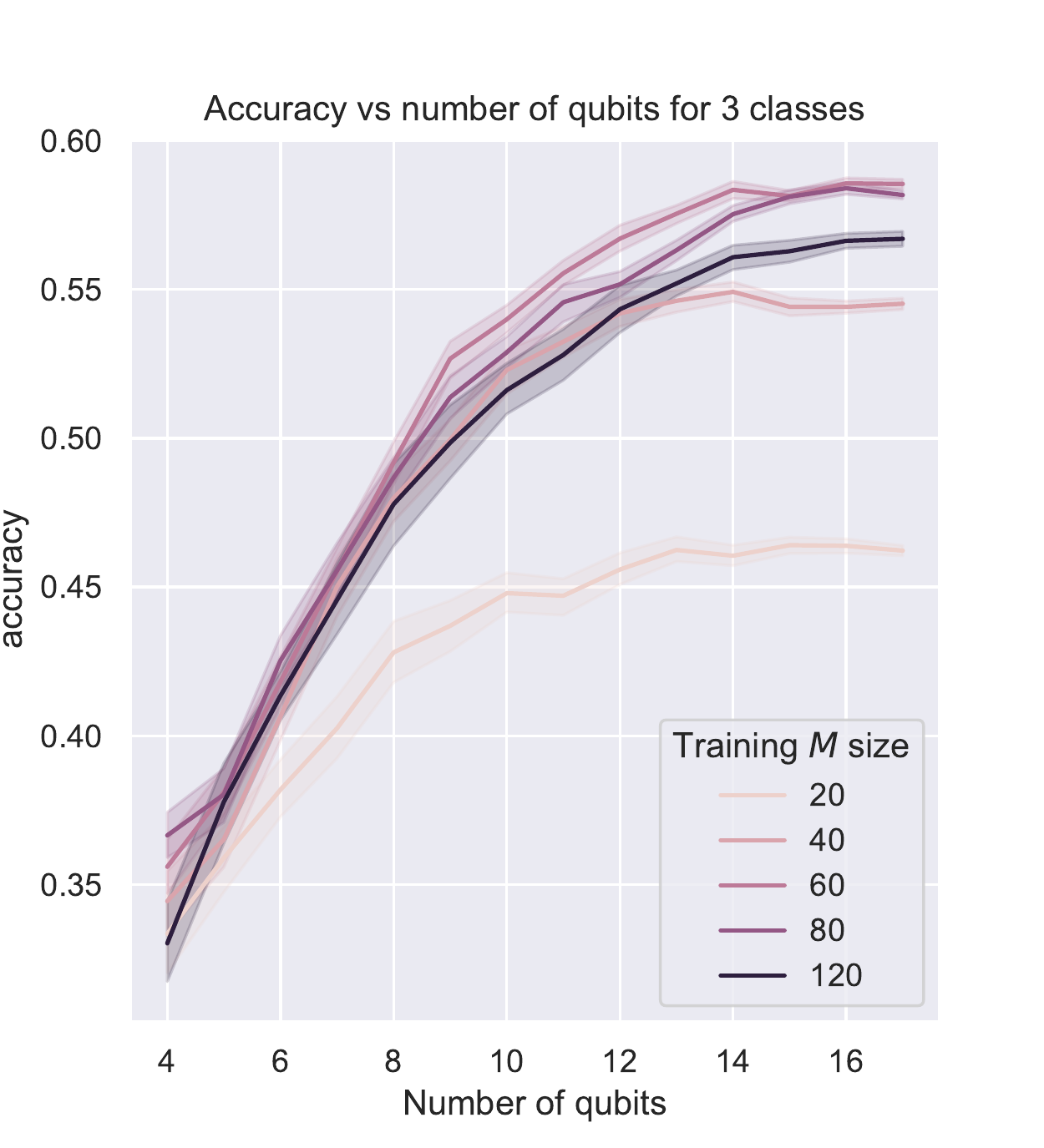}
    \caption{Accuracy vs number of qubits for 3 classes}\label{fig:acc3}
\end{figure}

\begin{figure}[!h]
    \centering
    \includegraphics[width=0.95\columnwidth]{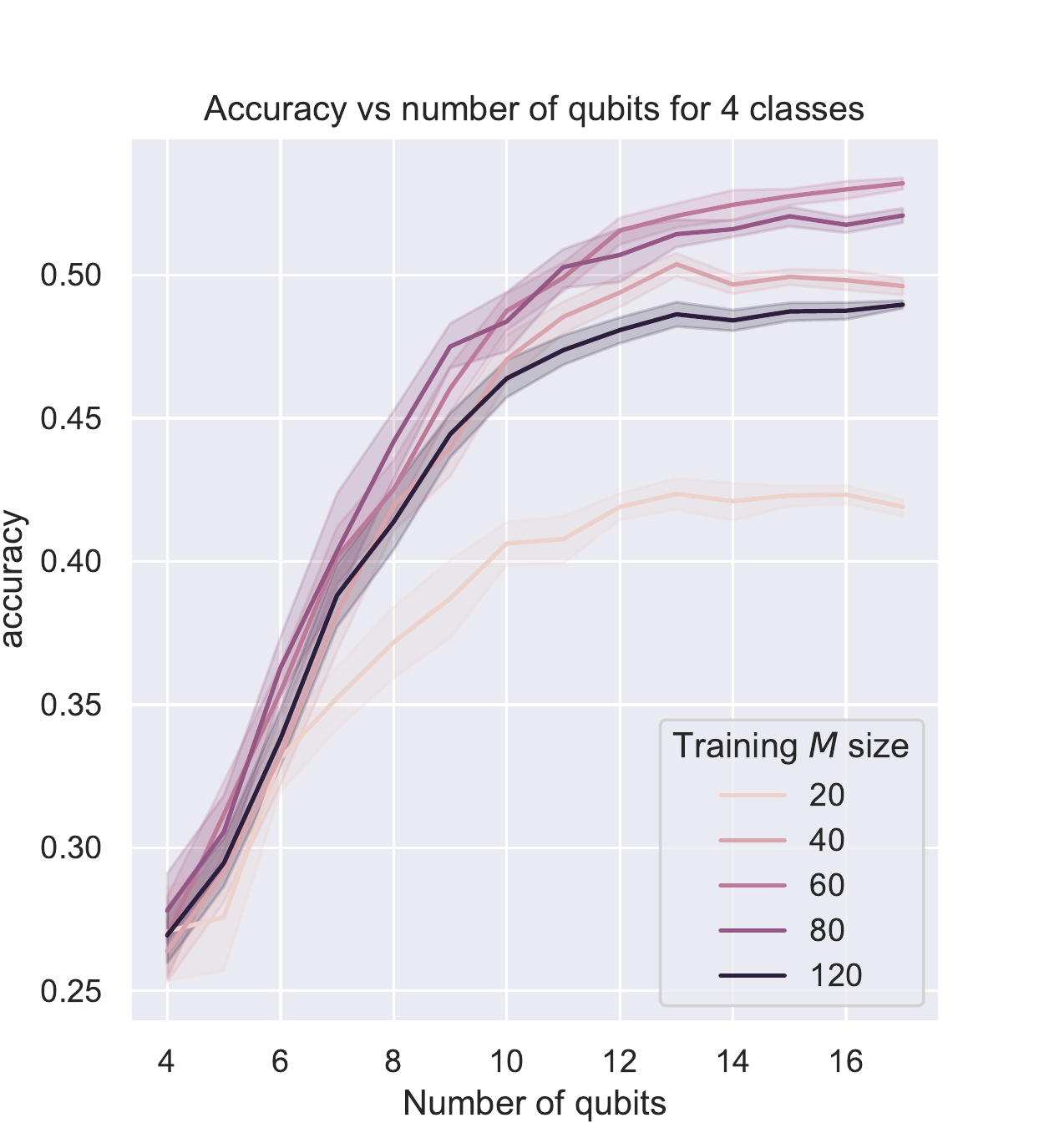}
    \caption{Accuracy vs number of qubits for 4 classes}\label{fig:acc4}
\end{figure}

\textbf{Results}
In Figures~\ref{fig:acc2},\ref{fig:acc3} and~\ref{fig:acc4} we report the accuracy
achieved for each simulation case.
We observe that
in all cases 
there is enough information extracted for meaningful video classification
even for as few as 10 qubits, for appropriate training sizes.
A quantum register of $q=10$ qubits encodes $2^{10} = 1024$ pixels, corresponding to just 
$2^{10} / (64\times64\times32) \approx 0.8\%$ of the initial video.
This is a significant reduction.
It appears that after approximately 
13 qubits no much more information is extracted.
We have to assume that this is at least partly due to our use
of a naive classification method
as one would expect that for a non-reduced dataset
(corresponding to $q=17$)
higher values of accuracy should be achievable.

The training size also plays a crucial role.
For training sizes of less than 40 videos per class
the classification accuracy was very low.
As the training size increased so did, in general, the classification accuracy.
Exceptions to this were observed as $k$ became larger and
we hypothesize this is due to the large similarities between the classes.
The classes (``Swiping Left'', ``Swiping Right'') and
(``Pulling Hand In'', ``Pushing Hand Away'')
are---in practice---time-reversed versions of each other.
On the other hand,
it is a strong point of the proposed method
that it can accurately discriminate between the time-reversed versions.




\section{Discussion}%
\label{sec:discuss}
The results obtained suggest
that 
a very small number of qubits
are able to contain enough information so that videos
can be successfully classified.
Even with the naive classification method
that we employed at the classification step,
the procedure achieved high accuracy values,
with a large data reduction 
and high performance.
Even at $q=5$ qubits there is a statistically significant deviation from random-choice classification
showing that meaningful information was extracted
using just 0.02\% of the initial video, corresponding
to a 4000 reduction of the video size.
The quantum method used is therefore
capable of accurately identifying the most significant
pixels using a very efficient procedure.

There are many ways that the above results
could be improved.
A better classification method, either classical or quantum,
is an obvious first step as the method we employed as 
proof of concept is very simplistic.
A more accurate extraction of the significant bits could also
be achieved by more detailed methods; for example, 
by using cross-validation in place of the simple video averaging.

Finally, as this method uses very few quantum operations and a few quantum bits,
it would be straightforward to implement it on a real quantum device
such as on an IBM qpu.

\textbf{Acknowledgment}
This  work  is  partially  supported  by  the  Greek  Secretariat  for  Research  and  Innovation  and  the  EU, Project  SignGuide:  Automated Museum Guidance using Sign Language T2EDK-00982  within  the  framework  of  “Competitiveness, Entrepreneurship and Innovation” (EPAnEK) Operational Programme 2014--2020.

\bibliographystyle{quantum}
\bibliography{q_data_reduction}

\end{document}